\documentclass[aps,english,prb,twocolumn,superscriptaddress]{revtex4}
\usepackage{graphicx}
\usepackage[left]{lineno}
\usepackage{gensymb}
\usepackage{amsmath}

\makeatletter
\def\p@figure{Fig.~}
\makeatother

\newcommand{\kappaph}{\kappa_\mathrm{ph}}
\newcommand{\tauel}{\tau_\mathrm{e}(E)}
\newcommand{\tauph}{\tau_\mathrm{ph}(\omega)}

\begin{document}

\title{Ultra-thin Carbon Biphenylene Network as an Anisotropic Thermoelectric Material with High Temperature Stability Under Mechanical Strain}

\author{G{\"o}zde {\"O}zbal Sargın}
\affiliation{Faculty of Engineering and Natural Sciences, Sabanci University, Istanbul, 34956 Turkey}

\author{Salih Demirci}
\affiliation{Department of Physics, Kırıkkale University, Kırıkkale, 71450 Turkey}

\author{Kai Gong}
\affiliation{Department of Civil and Environmental Engineering, Rice University, Texas, 77005 USA}

\author{V. Ongun \"{O}z\c{c}elik}\email{ongun.ozcelik@sabanciuniv.edu}
\affiliation{Faculty of Engineering and Natural Sciences, Sabanci University, Istanbul, 34956 Turkey}
\affiliation{Materials Science and Nano Engineering Program, Sabanci University, Istanbul, 34956 Turkey}

\begin{abstract}

Carbon biphenylene network (C-BPN), which is an ultra-thin material consisting of carbon atoms arranged  in square-hexagonal-octagonal (4-6-8) periodic rings, has intriguing properties for nano-scale device design due to its unique crystal structure. Here, using the Landauer formalism in combination with first-principles calculations, we show that C-BPN is a highly stable thermoelectric material at elevated temperatures under mechanical strain, where its thermoelectric efficiency can be anisotropically engineered. Transport calculations reveal that C-BPN's transmission spectrum has significant degrees of directional anisotropy and it undergoes a metal-insulator transition under strain, which leads to an increase in its Seebeck coefficient. C-BPN's lattice thermal conductance can be  selectively tuned up to 35$\%$ bidirectionally at room temperature by strain engineering. Enhancement in its power factor and the suppression of its lattice thermal conductance improves the p-type figure of merit up to 0.31 and 0.76 at 300 and 1000~K, respectively. Our findings reveal that C-BPN has high potency to be used in thermoelectric nano devices with selective anisotropic properties at elevated temperatures.
\end{abstract}

\maketitle

Designing thermoelectric (TE) materials that can efficiently convert energy between heat and electricity is one of the bottlenecks of solid-state energy harvesting problem where the ideal material should have high electrical conductivity and high Seebeck coefficient ($S$). In this field, nanomaterials and two-dimensional (2D) films are gaining interest due to their light weight, ease of production, flexibility, and variety of applications. Among these, 2D allotropes of carbon such as graphyne, penta-graphene, T-graphene, Net-graphene, and $\Psi$-graphene have been under spotlight of researchers since the successful realization of graphene.\cite{zhang2015penta,liu2012structural,li2017psi,ozcelik2013size,enyashin2011graphene} These pristine carbon-based materials and their functionalized forms have diverse periodic arrangements as well as symmetries in their crystals which lead to unique  physical properties such as ultra-high electron thermal conductivity, negative differential resistance, enhanced charge–discharge rate, superconductivity, high storage capacities for Na and Li-ion batteries. \cite{liu2012structural,tong2022ultrahigh,bravo2018tight,D3CP04143C,D2CP03155H,D1SE00657F}

Carbon biphenylene network (C-BPN), a two-dimensional (2D) material consisting of carbon atoms arranged  in 4-6-8 periodic rings, was  previously predicted based on first-principles calculations~\cite{wang2013prediction,denis2014stability}, and recently synthesized in 2021 by Fan \textit{et.al}\cite{fan2021biphenylene}. This material stands out due to its unique anisotropic $Pmmm$ symmetry (space group: No. 47) and its synthesis was followed by numerous discoveries on its vibrational, mechanical, magnetic and transport properties      \cite{demirci2022stability,mashhadzadeh2022theoretical,veeravenkata2021density,son2022magnetic,ying2022thermal}. Type-II Dirac fermionic and magnonic states of C-BPN's band structure offer novel topological features~\cite{liu2021type,son2022magnetic,zhang2023type}. Phonon analysis has shown that anisotropy in the mechanical properties of C-BPN may arise from inconsistent group velocities and different intensities of hybridization between two different types of carbon atoms in its structure.\cite{wang2022phonon} Based on classical molecular dynamics (MD) simulations, C-BPN has been predicted to have high melting point (4024~K) and Young's modulus (1019.4~GPa) which are comparable to the reported values for graphene.\cite{pereira2022mechanical} Altering the conductivity and mechanical properties of C-BPN has been achieved by strain engineering and doping.\cite{yang2023abnormal,ren2023tuning} Recently, Zhou \textit{et al.} showed that, buckled analogues of biphenylene can be promising candidates for nanomechanics and ferroelectric storage applications.\cite{doi:10.1021/acsanm.4c00454}
In addition to carbon, other group IV elements and group IV-IV, III-V and IV-VI compounds were also successfully integrated in the biphenylene geometry.\cite{demirci2022stability,demirci2022hydrogenated,gorkan2023can,
zhou2023predictions,guo2023lattice,ABDULLAHI2024113103,
doi:10.1021/acsanm.4c00699} 

Low-dimensional materials have an advantage on exhibiting high thermoelcetric performance due to the quantum confinement effect. Dresselhaus and Hicks have shown that confinement length can be used to optimize the electronic part of the figure of merit ($ZT$, a measures of the TE performance of materials) and reduce the lattice thermal conductance ($\kappa_{ph}$) value, \cite{hicks1993thermoelectric,hicks1993effect} which signals that C-BPN has a potential to be used as a thermoelectric material since low dimensional materials benefit from quantum confinement effect. However, pristine C-BPN also has a disadvantage due to its metallic character for utilization in TE applications, since an improvement in the TE figure of merit requires both high electrical conductivity and high $S$. Therefore, nano-engineering strategies such as cutting nanoribbons, constructing bilayers, introducing defects and alloying that can open the band gap of C-BPN are needed to increase the $S$ and thus  enhance the TE performance of C-BPN.\cite{xie2023intrinsic,lv2023funnel,chowdhury2022first}
 
\begin{figure*}
\includegraphics[width=145mm]{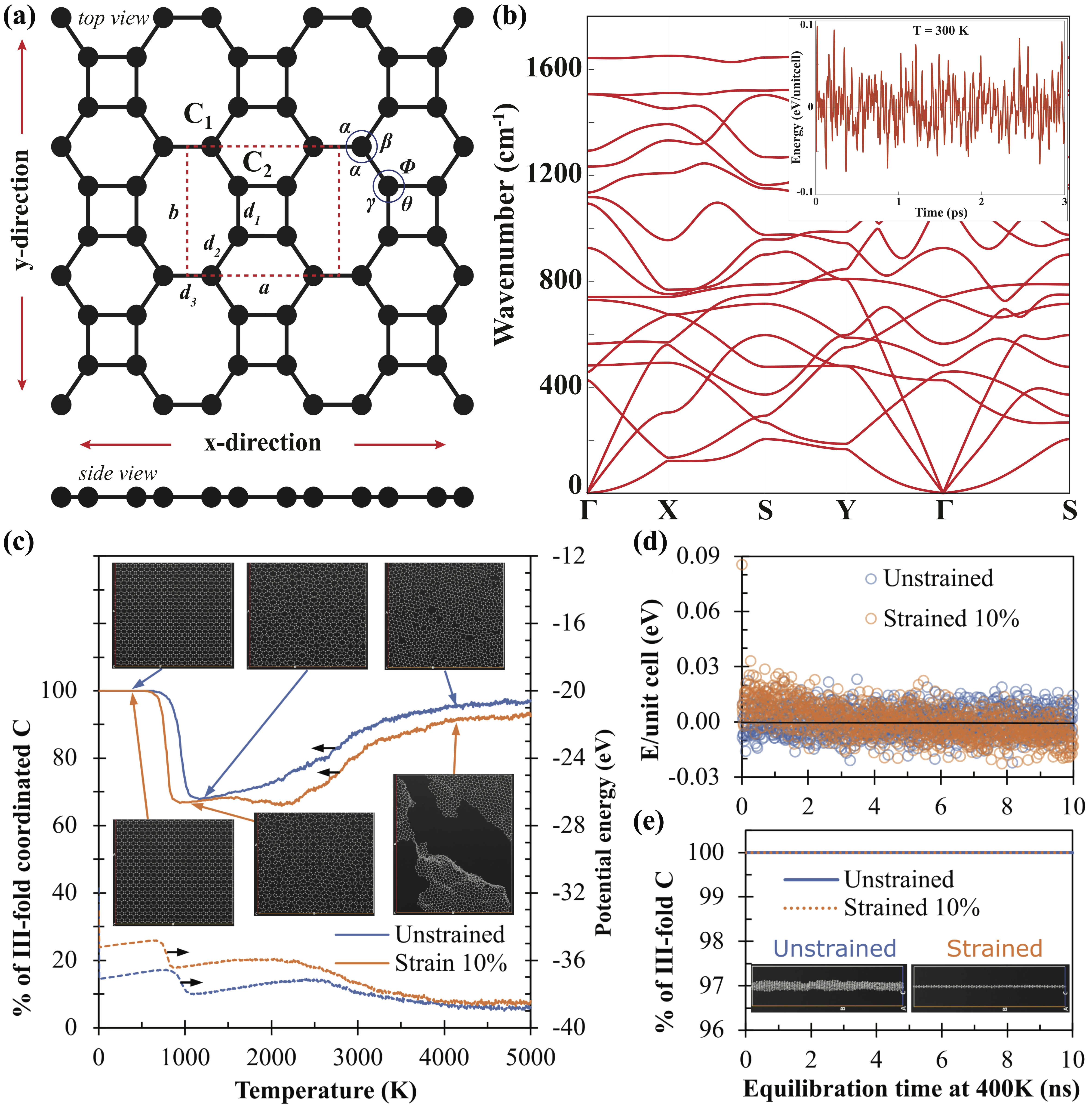}
 \caption{(a) Top and side views of the C-BPN monolayer. Its unit cell is indicated by the dashed rectangle with lattice constants $a$=4.52\AA~ and $b$=3.76\AA~. The structure has different C-C bond distances with $d_1$=1.46\AA~, $d_2$=1.41\AA~, $d_3$=1.44\AA~ and bond angles with $\alpha$=125\textdegree, $\beta$=110\textdegree, $\gamma$=145\textdegree, $\phi$=125\textdegree,  and $\theta$=90\textdegree. (b) The phonon dispersion relation of C-BPN under zero strain. The inset shows the fluctuation of C-BPN's the total energy calculated by AIMD simulations at room temperature. (c) The evolution in the percentage of III-fold coordinated carbon atoms (left axis) and potential energy (right axis) in both unstrained and strained C-BPN as the temperature increases from 1 to 5000 K over 10 ns. The inset figures show the C-BPN structures at selected temperatures. (d) The fluctuations in the total energy (in eV) per unit cell, and (e) the percentage of III-fold coordinated carbon atoms for both unstrained and strained C-BPN during 10 ns of equilibration at 400 K. The inset figures in (e) show the side views of the unstrained and strained C-BPN structures.}
 \label{fig1}
\end{figure*}

Here, we show that monolayer C-BPN  maintains its structural stability at elevated temperatures under mechanical stress and its TE efficiency can be selectively engineered biaxially (in the x-and y-directions) by strain engineering. We conduct ballistic transport calculations of stable C-BPN structures using the Landauer formalism combined with \textit{ab-initio} methods based on density functional theory (DFT).\cite{SupplementaryMaterial} Band gap opening is accomplished at 10$\%$-biaxial tensile strain and we show that direction-dependent geometry manifests itself in both the electronic and the phonon transmission spectra. Improvement in the $S$ of C-BPN was  achieved due to band gap opening and the emergence of nearly flat bands in its electronic structure under strain. Our results also show that strain engineering at elevated temperatures can enhance the electronic $TE$ coefficient of C-BPN, as well as allowing the selective engineering of its lattice thermal conductance anisotropicly  such that, it can be tuned up to 35$\%$ along the x- and y-directions at room temperature.
 
\begin{figure*}
 \centering
\includegraphics[width=15cm]{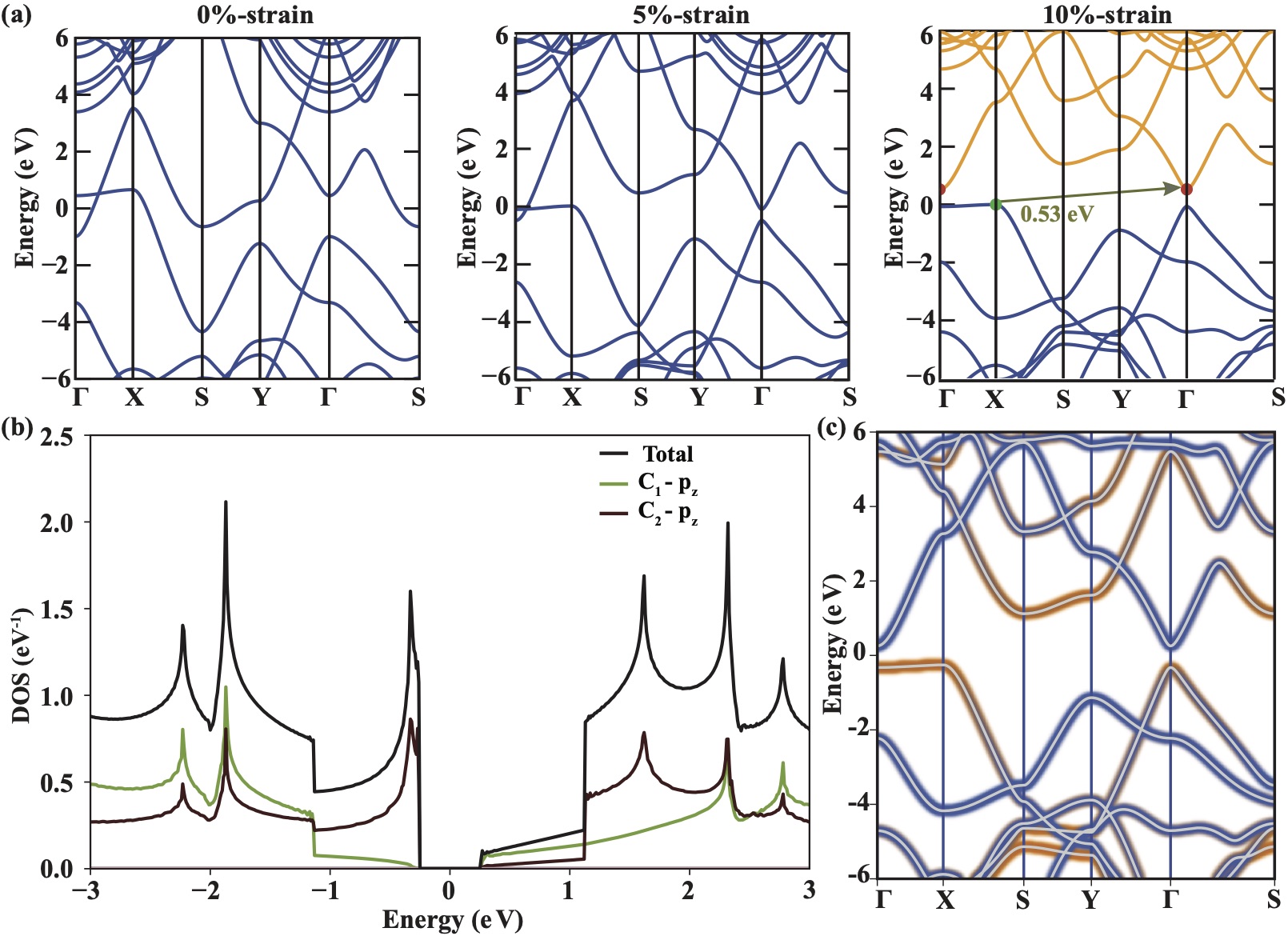}
 \caption{(a) Electronic band structures of C-BPN calculated using the HSE06 functional from 0$\%$ to 10$\%$ biaxial tensile strain, (b) Total and site projected DOS, (c) Site projected electronic band structure (Yellow color represents influence of the C$_2$ atoms while blue color denotes weight of C$_1$ atoms.) obtained with HSE06 functional under 10$\%$ biaxial tensile strain. }
 \label{fig2}
\end{figure*}

\subparagraph{Structural stability of C-BPN:}
The structural properties of pristine C-BPN and its phonon diagram optimized by DFT are shown in \ref{fig1}. The phonon diagram indicates stability since all vibrations have positive frequencies.  Room temperature stability of C-BPN is tested by calculating the fluctuations of total energy using \textit{ab-initio} MD simulations, which show that the amplitude of oscillation in total energy (0.08eV) is much lower than the cohesive energy of the material (7.50eV).\cite{demirci2022stability}  High-temperature stability tests using a reactive force field MD was used to further justify the material's structural stability at elevated temperatures and under external mechanical stress. \ref{fig1}(c) illustrates the evolution of the percentage of III-fold coordinated carbon atoms in both unstrained and strained C-BPN as they are heated from 1 to 5000 K at a rate of 0.5 K/ns, employing a force field MD approach. The unstrained and strained C-BPN exhibit 100\% III-fold coordination below a temperature of approximately 540 and 640 K, respectively, indicating stability. The lower threshold for the strained structure suggests that its thermal stability is slightly lower than the unstrained case. Beyond these thresholds, a discernible emergence of non-III-fold coordination in carbon atoms is observed, suggesting the occurrence of C-C bond breaking and formation. This phenomenon is visually depicted in the inset of \ref{fig1}(c), which show the breaking of C squares and the formation of holes, particularly evident in the case of strained C-BPN. 

Following a period of decrease, the percentage of III-fold coordinated carbon atoms begins to rise with increasing temperature. Concurrently, there is a heightened formation of hexagonal carbon structures, as evidenced by the angular distribution evolution.\cite{SupplementaryMaterial} The transformation from C squares and pentagons to C hexagons during this phase is associated with a reduction in potential energy, indicating the energetic favorability of this transformation. To further assess the stability of unstrained and strained C-BPN, we subjected them to a 10 ns heating process at 400K using force field MD. In \ref{fig1}(d), the fluctuations in the total energy per unit cell for both C-BPN structures oscillate around an average value with an amplitude of 0.03eV per unit cell. This amplitude is significantly smaller than the cohesive energy of the structure (7.50eV), suggesting that these oscillations are insufficient to break apart the monolayer. Coordination analysis in \ref{fig1}(e) reveals that both unstrained and strained C-BPN maintains 100 \% III-fold coordination through the heating process indicating that both of them are stable at 400K.  The inset figures in \ref{fig1}(e) show the straight structure of strained C-BPN, whereas unstrained C-BPN exhibits surface wrinkles.

\begin{figure*}
 \centering
\includegraphics[width=15cm]{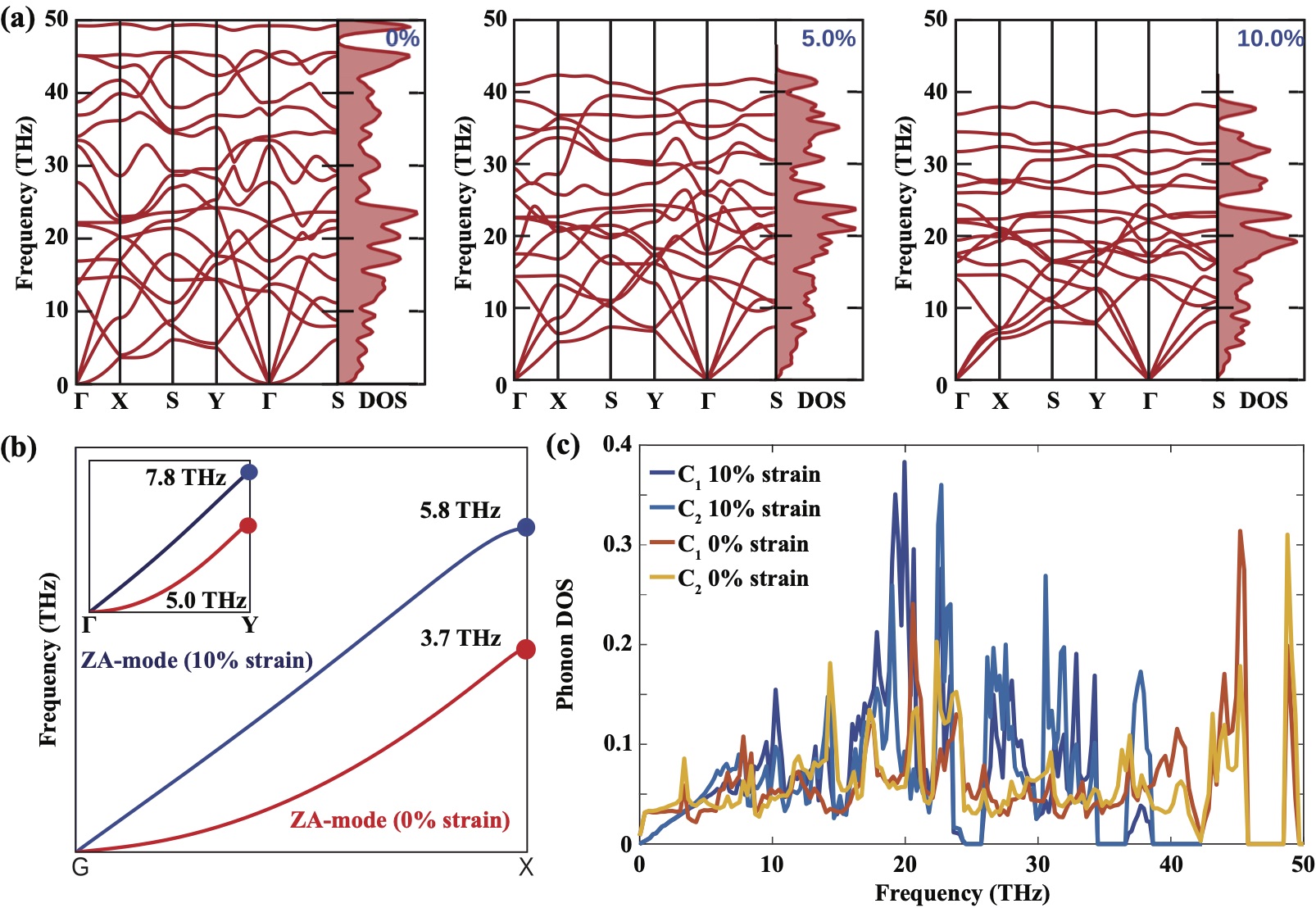}
 \caption{(a) Phonon dispersion relations with the total phonon density of states of unstrained and strained C-BPNs, (b)  ZA modes along x and y direction of unstrained and strained C-BPN (10$\%$) and (c) phonon projected DOS.}
 \label{fig3}
\end{figure*}

\subparagraph{Anisotropic evolution of the electronic properties:}

Having justified the high temperature stability of C-BPN, we next focus on the evolution of its electronic structure under strain. C-BPN shows an intriguing electronic band structure which possesses a type-II Dirac cone along the $\Gamma-X$ (x-direction) path as  shown in \ref{fig2}(a). Here, type-II Dirac point is originated from two crossed bands in which one of the crossing bands is nearly flat and the other one is highly dispersive. Previous studies have shown that, this unique band structure type is present in topological states, type-II Dirac magnons and superconductors.\cite{liu2021type,zhang2023type,liu2023superconductivity} As the strain value reaches 10{\%}, the material transforms from a metal to a semiconductor with an indirect band gap of 0.53eV based on DFT calculations using HSE06\cite{heyd2003hybrid,krukau2006influence} functional (0.11eV based on PBE\cite{perdew1996generalized}  calculations)  \cite{SupplementaryMaterial} by opening the tilted Dirac cone along the $\Gamma-X$ path. 

The effective masses of charge carriers are calculated using the inverse of the local curvature of the electronic band energy by $m^{*}=\hbar \left(\frac{\partial E^2 (\mathbf{k})}{\partial k^2}\right)$. To calculate the derivative of the band band energy  with respect to the k, the finite difference method was used instead of simple parabolic fit in order to obtain a more accurate $m^*$. The effective mass of holes along $\Gamma$-X (x-direction) and $\Gamma$-Y (y-direction)  are highly anisotropic. The nearly flat valence band along the x-direction results in $m^{*}=8.02m_0$, while the linear-like valence band (VB) along $\Gamma$-Y gives $m^{*}=0.03m_0$, where $m_0$ is the free electron mass. In the lowest conduction band (CB), the parabolic portion has $m^{*}=0.194m_0$ along the x-direction and the linear-like portion induces $m^{*}=0.03m_0$ along the y-direction. \ref{fig2}(b) shows that nearly flat portion of the VB gives rise to sharp peak in total density of states (DOS) at the VB edge. This means that more conducting channels will be allowed to participate in the p-type electronic transmission. The main contribution to the total DOS comes from $p_z$ orbitals of the C$_2$ type carbon atoms, while the remaining contribution originated from $p_z$ orbitals of C$_1$ type atoms at the VB edge. Contrarily, at the CB edge, total DOS is smaller than it is at the VB edge as is evident from having more dispersive band shape compared to the VB. Unlike the VB, contribution of the C$_1$ atom is larger than that of C$_2$ at the CB edge. In addition, the site projected electronic band structure of strained C-BPN, illustrated in \ref{fig2}(c), confirms that the valence band maximum and the conduction band minimum are dominated by  C$_2$ and C$_1$ atoms, respectively.

\subparagraph{Phonons and lattice thermal conductivity:}
Phonon dispersion relations, which are important for  dynamical stability and lattice thermal conductivity, are presented in \ref{fig3}(a) for pristine and strained C-BPNs. There are no imaginary frequencies in the phonon spectra of  pristine  and biaxially strained C-BPN.  The maximum phonon frequency of unstrained C-BPN  reaches to 49.5 THz which is close to that of graphene.\cite{VEERAVENKATA2021893}  As the biaxial tensile strain increases, the quadratic nature of the ZA mode around the $\Gamma$ point turns into linear along the x- and y-direction, where the group velocity ($\upsilon_{g,s}(\boldsymbol{q})=\partial \omega_{\alpha}(\boldsymbol{q})/\partial \boldsymbol{q}$) of the linearized ZA mode is still smaller than that of the TA and LA modes. At zero strain, ZA phonon mode follows $q^2$ dependency up to 3.7THz and  5.8THz  for $\Gamma$-X and $\Gamma$-Y directions, respectively,  which are the highest frequencies of zone boundaries along these directions, as presented in \ref{fig3}(b). This leads the phonon transmission function of C-BPN along the $\Gamma$-Y direction to be larger than it is along the $\Gamma$-X direction at low energies. The group velocities of LA and TA modes along the $\Gamma$-X ($\Gamma$-Y) direction are 18.505km/s (19.801 km/s)  and 11.380km/s (11.372km/s), respectively,  as $q \rightarrow 0$. However at the BZ boundary, phonon anisotropy becomes more prominent.\cite{wang2022phonon} Therefore, it is expected that $\tauph$ and $\kappaph$ of C-BPN to take higher values along the $\Gamma$-Y direction than the $\Gamma$-X direction. As presented in \ref{fig3}(a), phonon frequencies shift down when the interaction between the atoms are weakened due to external strain. The highest phonon frequency reduces to 38.5 THz when the applied strain reaches to 10$\%$. 

According to the site projected phonon DOS diagram,  C$_1$ and C$_2$ atoms contribute less to the phonon densities for the strained system than the unstrained case at lower frequencies, which is due to the linearization of the ZA mode under strain as shown in \ref{fig3} (c). However, phonon DOS of the strained system becomes dominant at  moderate frequencies since the phonon bands become flattened with increasing strain. Phonon DOS of the strained C-BPN goes to zero at low frequencies as also seen from the opening of the phonon band gap between optical phonon modes. Enhancement in the DOS of strained C-BPN contributes to phonon transmission spectrum while decrease in group velocities of LA and TA modes have detrimental impact on phonon transmission spectrum. Since the $\tauph$ depends on the product $\delta(\omega-\omega_{\alpha}(q))\nu_{\alpha}(q)|_{\parallel}$~\cite{karamitaheri2013ballistic,datta2005quantum}, competition between the phonon DOS and the group velocity determines the trend of the phonon transmission spectrum.

\begin{figure*}
\includegraphics[width=15cm]{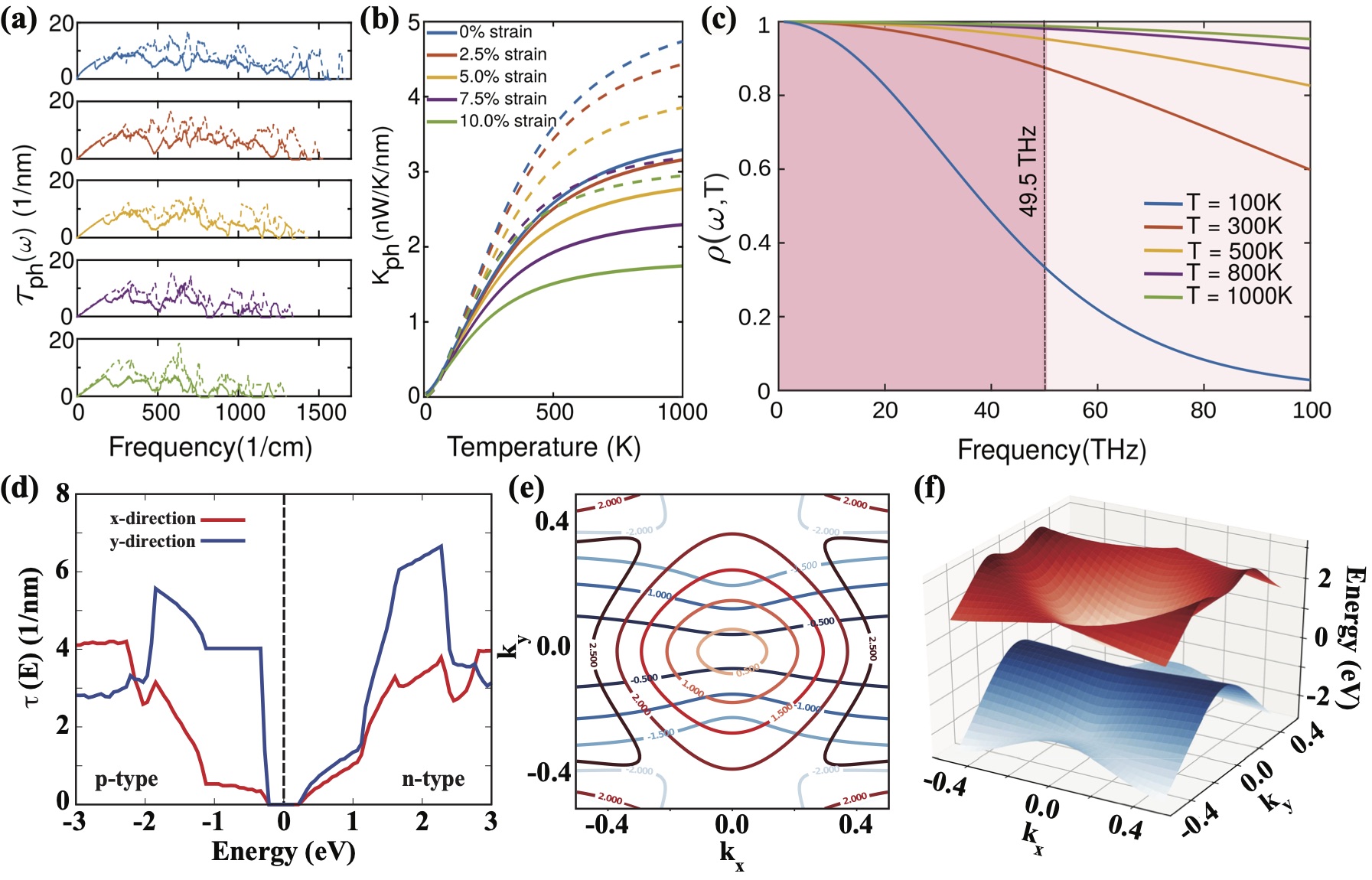}
 \caption{(a-b) Phonon transmission spectra ($\tauph$) and the lattice thermal conductance ($\kappa_{ph}$) along the x-direction (solid curves) and y-directions (dashed curves) for 0$\%$ to 10$\%$ biaxial tensile strain. (c) Phonon window function as a function of frequency for various temperatures ranging from 100 K to 1000 K. Maximum phonon frequency of the unstrained C-BPN is indicated with the dashed vertical line. (d) Normalized electronic transmission spectrum (by width in nm), (e) isoenergy surfaces for the two uppermost VB and lowermost CB and (f) the three dimensional visualization of the topmost VB and the lowest CB. The results are obtained using HSE06 under $10\%$ strain along the x- and y-direction.}
 \label{fig4}
\end{figure*}

We also observe anisotropy in the phonon transmission spectrum($\tauph$) of C-BPN as shown in \ref{fig4}(a). Phonon transmission decreases with external strain and it gets squeezed into a narrower energy range in line with the phonon dispersion. As strain increases, fewer conducting channels contribute to phonon transport. Due to the higher frequencies of acoustic phonons along the y-direction, $\tauph$ reaches higher values along the y-direction than the x-direction for all applied strain values. $\kappa_{ph}$ is calculated by integrating the product of phonon window function $\rho(\omega,T)$ and $\tauph$ over all phonon frequencies. Phonon window function, $\rho(\omega,T)=\hbar\omega(\partial f_{BE}(\omega,T)/\partial T)$, determines the contribution of phonon modes to the $\kappaph$. 

The dependence of $\rho(\omega,T)$ to  frequency and temperature can be seen in \ref{fig4}(c). As temperature increases, the  width $\rho(\omega,T)$ enlarges and approaches to 1 even for low frequencies. At low temperatures $\rho(\omega,T)$ filters out the high frequency optical modes, and acoustic modes dominate the $\kappaph$. However for low strain values at 1000~K,  $\rho(\omega,T)$ converges to 98$\%$ of its maximum value which will be achieved with the contribution of optical modes above 1000~K. When there is no strain in the y-direction, $\kappa_{ph}$ has its maximum value while its minimum value appears when there is 10$\%$ strain in the x-direction. With regard to TE efficiency, low lattice thermal conductance is favorable. Since at high temperatures $\rho(\omega,T)$ approaches to 1, $\kappaph$ starts to saturate. The $\kappaph$ values are calculated as 1.76 and 2.38$nW/(Knm^{-1})$ along the x- and y-directions under zero strain at room temperature. Under biaxial tensile strain,  $\kappaph$ decreases to two-third of its unstrained value along the x-direction. The influence of the biaxial tensile strain on $\kappaph$ along the y-direction is relatively weak as compared to the x-direction. The $\kappaph$   values drop to 1.14 and 1.90 along the x- and y-directions under 10$\%$ strain at 300~K. The $\kappaph$ values of C-BPN are comparable to those reported for zigzag and armchair graphene nanoribbons (1.64 and 1.15$nW/K$). \cite{xie2023intrinsic}

\subparagraph{Strain dependent electronic transmission:}
Strain-dependent electronic transmission spectra, $\tauel$, of C-BPNs under biaxial tensile strain ranging from 0$\%$ to 10$\%$  were calculated using their electronic band diagrams (obtained by HSE functionals) where the  $\tauel$ values were normalized according to the width of their transport channels (see Figure~S4).\cite{SupplementaryMaterial}. Pristine C-BPN has direction-dependent electron transmission character where the transmission is higher in the y-direction in comparison to the x-direction.~\cite{alcon2022unveiling} It should be noted that, due to the existence of available transport channels around its Fermi level, the metallic C-BPN under zero strain has the highest transmission value in comparison to C-BPNs under strain. 

When the applied strain reaches  10$\%$,  $\tauel$ drops to zero both in the x and y directions around the Fermi level as shown in \ref{fig4} (d). In this case, the n-type $\tauel$ is  higher than the p-type $\tauel$ in the x-direction, while the p-type $\tauel$ is significantly higher than the n-type $\tauel$ along the y-direction.   As shown in \ref{fig4} (d), the p-type $\tauel$ steeply increases at the VB edge along the y-direction, which is due to the existence of the nearly flat band along the $\Gamma-X$ direction in the electronic band diagram of C-BPN under 10$\%$ biaxial strain (\ref{fig2} (a)). The parabolic CB creates a square root energy dependence of the n-type $\tauel$ along the y-direction, whereas the linear portion of the CB creates a linear dependence along x-direction. Therefore, the transmission spectra also demonstrate that there is a strong anisotropy in the electronic transport properties of strained C-BPNs. The energy isosurface plots presented in \ref{fig4} (e) also show the directional contribution of the conduction and valence bands to $\tauel$. When the Fermi level crosses the topmost VB, cigar-shaped isosurfaces elongate in the x-direction. As we go down into deeper energy states, isosurfaces get wider along the y-direction and the cigar-shaped isosurfaces disappear, which indicate that  p-type $\tauel$ along y-direction is higher than p-type $\tauel$ along x-direction. The three-dimensional visualizations of VBM and CBM  presented in \ref{fig4} (f) also shows the strong anisotropy discussed above.

\begin{table}[t]
\caption{\label{tab:literature_ZT}
\label{tab_kappa_ph}
Reported $ZT$ and $\kappaph$ values of 2D carbon-based materials at room temperature.}
\begin{tabular}{cccc}\\
\hline
2D-Material&ZT(p/n)&$\kappaph$ \\
&&($Wm^{-1}K^{-1}$)\\
\hline
Graphene~\cite{reshak2014thermoelectric}&0.08&14.1\\
&\textsuperscript{\emph{a}}0.56&-\\
Graphene~\cite{anno2017enhancement}&0.55$\times$ 10$^{-3}$&3367.6$\pm$ 485.0\\
&\textsuperscript{\emph{b}}2.65&-\\
Graphene~\cite{wang2013thermoelectric}&-/0.0094&-\\
Graphyne~\cite{wang2013thermoelectric}&-/0.157&- \\
$\gamma$-Graphyne~\cite{jiang2017thermoelectric}&0.49/0.59(x)&76.4\\
&0.48/0.56(y)&-\\
\textsuperscript{\emph{c}}Phagraphene\cite{ghosh2022optical}& 0.58&0.58 \\
CBN~\cite{lv2023funnel}&1.80/-(armchair)& 33.19 \\
Graphdiyne~\cite{tan2015high}&0.99/3.26 &22.3\\
$\gamma$-Graphyne~\cite{tan2015high}&0.91/1.81&82.3 \\
Penta-graphene~\cite{chen2017enhanced}&0.0531/0.019&197.85 \\
\textsuperscript{\emph{d}}0.481/0.148&116.73\\
Graphenylene~\cite{zhang2023thermal}&-/0.48(x)&378\\
&-/0.39(y)&-\\
\textsuperscript{\emph{e}}C-BPN&0.26$\times$10$^{-2}$/0.24$\times$10$^{-3}$(x)&1.76\\
&0.47$\times$ 10$^{-2}$/0.21$\times$ 10$^{-4}(y)$&2.38\\
\textsuperscript{\emph{f}}C-BPN&0.06/0.04(x) &1.14\\
&0.31/0.04(y)&1.90\\
\hline
\end{tabular}\\
\textsuperscript{\emph{a}}H$_2$S adsorption
\textsuperscript{\emph{b}}Defect-ZT/ZT$_0$
\textsuperscript{\emph{c}}Electronic $ZT$
\textsuperscript{\emph{d}}Strained
\textsuperscript{\emph{e}}Our work-unstrained
\textsuperscript{\emph{f}}Our work-strained
\label{tab:ZT_table}
\end{table}

\begin{figure}
 \centering
\includegraphics[width=10cm]{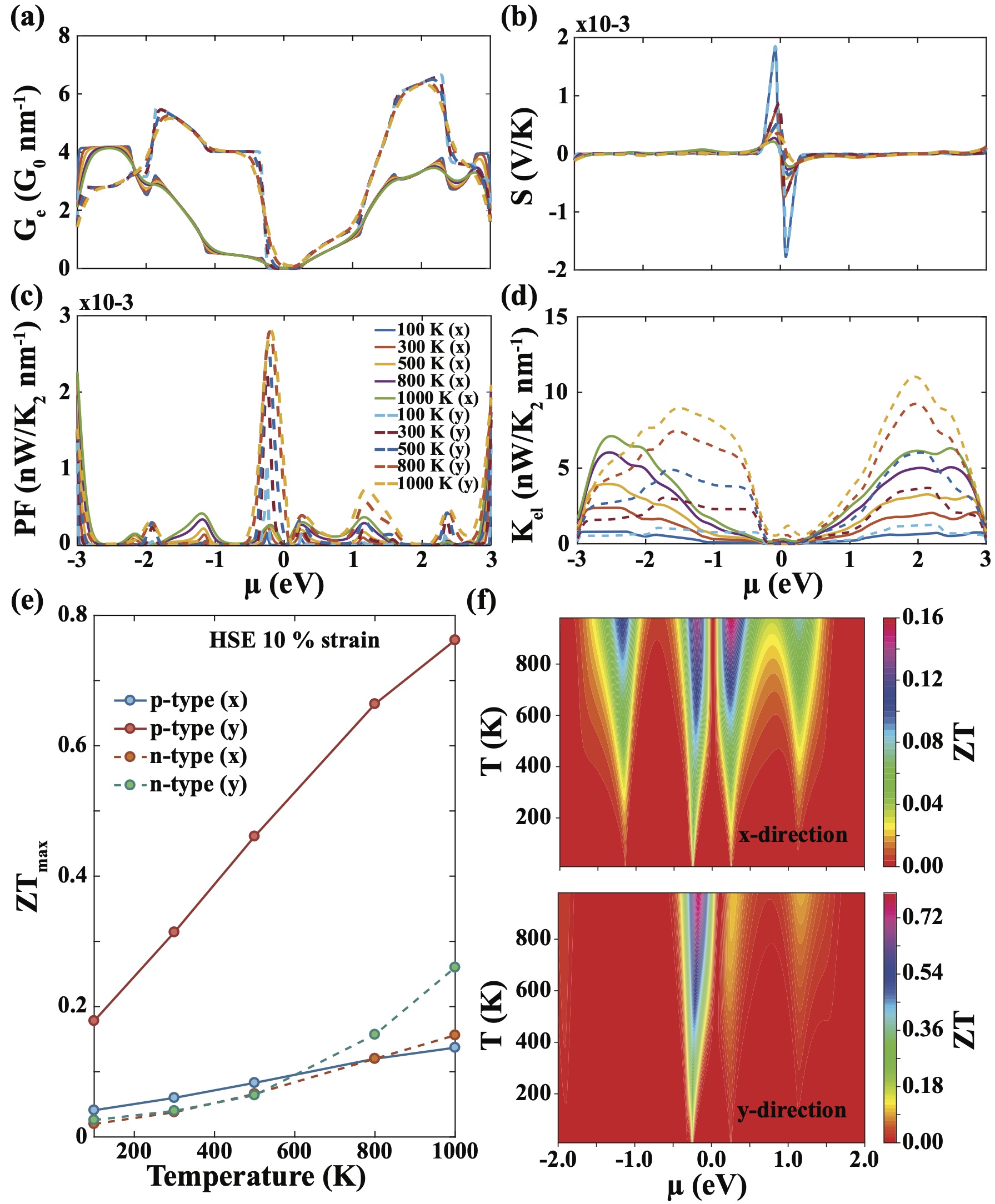}
 \caption{(a) Electronic conductance, (b) Seebeck coefficient, (c) power factor, (d) electronic thermal conductance of C-BPN from 100K to 1000K. (e) The maximum p- and n- type $ZT$ values as a function of temperature and (f) $ZT$ values as a function of temperature and chemical potential along x- and y-direction under $10\%$ biaxial strain.}
 \label{fig5}
\end{figure}

\subparagraph{Anisotropic thermoelectric transport:}
Having seen anisotropy in the electronic transmission properties of C-BPNs, we next focus on the anistropic TE transport properties of C-BPN and investigate the effects of strain on the $ZT$. The electronic TE properties can be derived using the $L_n$ integrals, as detailed in the Methods section. In ~\ref{fig5} (a), the electrical conductance, $G_e$, as a function of the chemical potential, $\mu$, for C-BPN under 10$\%$ strain is presented in terms of the quantum conductance unit ($G_0 = 2e^2/h$)  calculated at the HSE level.  In the limit $T \rightarrow 0$, $G_e$ approaches to $\tauel$ while it softens with increasing temperature. A strong anisotropy in the p-type $G_e$ along the x-and the y-direction is clearly seen near the Fermi level. Unlike the p-type $G_e$, n-type $G_e$ exhibits almost the same trend up to 1eV. Above 1eV, the anisotropic behavior appears with the contribution of the second CB to the transport. In contrast to the high $G_e$,  $S$  takes its lowest values in the metallic C-BPN, resulting from the bipolar conduction (shown in  Figure~S6\cite{SupplementaryMaterial}). The increase in the $S$ of C-BPN under 10$\%$-strain is due to the band gap opening (up to 0.53eV), which is sufficient to avoid bipolar conduction effect up to 600~K. Above 600~K, $S$ starts to decrease due to the bipolar effect. While the increase in the power factor ($PF$) and the $ZT$ accelerate with temperature, it slows down due to the decrease in the the $S$. Improvement of p-type $S$ along the y-direction is not only due to the band gap opening but also results from the step-like increase in the $\tauel$ at the the VB edge which is evident from the Mott's relation\cite{cutler1969observation,paulsson2003thermoelectric}  given by  $S(T,\mu)=-(\pi^2 k_{B}^2 T/3|e|)$[$\partial$ln$\tau(E)$/$\partial E$]. Accordingly,  as the slope of  $\tauel$ increases,  $S$ should also increase. Combining $G_e$ and $S$ gives the power factor, which plays a key role in determining TE performance. The p-type $PF$ along the y-direction is significantly higher than it is in the x-direction. The electronic thermal conductance follows the same trend with  $G_e$.

The $ZT_{max}$ values as a function of temperature for C-BPN at 10$\%$ strain calculated at the HSE06 level of theory is presented in \ref{fig5}(e). (Results at the PBE level of theory are available in the Supporting Information.) The fact that the p-type $ZT_{max}$ along the y-direction is much higher than it is along the x-direction is a another result of the anisotropy of C-BPN, which supports the above discussions. Table~\ref{tab:ZT_table} summarizes the maximum $ZT$ as well as $\kappaph$ values of known 2D carbon-based materials and our work at for 300K. Here $\kappaph$ indicates lattice thermal conductance, normalized conductance or conductivity depending on the transport regime. Accordingly, the TE performance of C-BPN is larger than similar carbon-based monolayer semiconductors, such as  graphyne, which is a direct semiconductor.\cite{wang2013thermoelectric} The p-type $ZT_{max}$ along the y-direction is 0.31 at 300K which is much higher than TE efficiency of graphene as seen from \ref{fig5}(e).  Moreover, $ZT$ of strained C-BPN is comparable to those of $\gamma$-graphyne, phagraphene, graphenylene and penta-graphene with induced strain as presented in Table~\ref{tab:ZT_table}. The calculated p-type  $ZT_{max}$ along the y-direction reaches 0.76 at 1000~K. The p- and n-type $ZT_{max}$ along the x-direction achieves values around 0.15 at 1000~K, while n-type $ZT_{max}$ along the y-direction shows significant increase above 600~K and reaches 0.26. In order to provide a broader perspective into the TE efficiency of C-BPN for both directions under 10$\%$-strain, $ZT$ is plotted as a function of temperature and chemical potential in \ref{fig5}(f). One can see that the difference between the p- and n-type $ZT$ in the y-direction is significantly large, while the n- and p-type efficiencies in the x-direction are close to each other.

In conclusion, using the Landuer formalisim in combination with DFT calculations, we reveal that C-BPN is a highly stable material even at elevated temperatures and external mechanical strain where its TE properties can be selectively engineered bidirectionally using strain engineering.  The material undergoes a  metal-insulator transition at elevated strain values where the band gap opening improves its $S$ and power factor leading to improved TE performance under strain. In addition, we show that flat band portion of the valence band gives rise to an increase in p-type electronic TE coefficients of C-BPN. Dissimilar effective masses of holes along the x- and y-directions cause a strong anisotropy in the DOS, electronic transmission spectra, and hence electronic TE coefficients. Direction-dependent anisotropy also exists in the phonon dispersion of C-BPN which induces distinct $\kappa_{ph}$ values for x- and y-direction. Lattice thermal conductance $\kappa_{ph}$ is suppressed by 35$\%$ along x-direction while it is reduced by 20$\%$ along y-direction at 300~K via strain engineering compared to its unstrained values. The combined effect of the enhancement in its power factor and the suppression of its $\kappa_{ph}$ increases the p-type $ZT$ along y-direction up to 0.31 and 0.76 at 300 and 1000K, respectively.  Our findings reveal that ultra-thin carbon biphenylene network is stable and has high potential to be used as a TE material even at elevated temperatures and external mechanical stress.

\clearpage

\subparagraph{Methods:}

Density functional theory  calculations were carried out using the Vienna \textit{ab-initio} simulation package (VASP).\cite{hafner1997vienna,kresse1996efficient} The projector augmented wave (PAW) method was employed to define the 
the pseudopotential of Carbon atom.\cite{blochl1994projector,kresse1999ultrasoft} The exchange-correlation potential has been treated by generalized gradient approximation (GGA) using Perdew–Burke–Ernzerhof (PBE) functionals.\cite{perdew1996generalized} Energy cutoff value for plane wave basis set was taken as 520eV. Brillouin zone (BZ) is sampled by a $\Gamma$-centered 12 × 14 × 1 $k$-meshes.\cite{monkhorst1976special} Simulation cell was adopted with a vacuum spacing of 20Å along the z direction to eliminate the artificial interactions between adjacent monolayers. During the structural relaxation, the convergence criteria for energy difference between electronic steps was set to $10^{-5}$ eV and the force on each atom  was set to$10^{-2}$eV/Å. Since GGA PBE formalism underestimates the electronic band gaps, electronic band
structures were also obtained by applying Heyd–Scuseria–Ernzerhof (HSE06) functionals method.\cite{heyd2003hybrid,krukau2006influence} Dynamical stability analysis were performed by density functional perturbation theory (DFPT) \cite{baroni2001phonons} as implemented in the PHONOPY package.\cite{togo2015first,togo2023first,togo2023implementation} In phonon dispersion calculations, supercells were taken sufficiently large 4 × 5 × 1. Electronic and phonon band structures are plotted with SUMO toolkit.\cite{ganose2018sumo}

The thermal stability of the systems was initially tested by \textit{ab-initio} MD simulations using a microcanonical ensemble method (NVT)~\cite{nose1984unified,hoover1985canonical} with constant temperatures (300~K), with time steps of 1 fs and a total simulation time of 3 ps. The stability of both unstrained and strained C-BPN structures was further evaluated using force field MD simulations, facilitated by the QuantumATK NanoLab software package.~\cite{Schneider_2017,smidstrup2020}  These simulations employed a three-body Tersoff potential,~\cite{tersoff1988new,tersoff1988empirical} which has been widely used for modeling carbon-based materials such as graphene and carbon nanotubes. This potential explicitly considers the bond order contingent upon the local atomic environment and includes a short-range interaction term delineated by the universal Ziegler-Biersack-Littmark (ZBL) potential.~\cite{ziegler1985stopping}  The parameters for the force field were taken from previous the literature.\cite{bellido2012molecular} Each structure, consisting of a 20x20 supercell with 2400 atoms, was subjected to an MD run spanning 107 steps, implemented under periodic boundary conditions within an NVT ensemble and controlled by the Nose–Hoover thermostat.\cite{martyna1992nose} A timestep of 1fs was employed, culminating in a total simulation time of 10ns for each scenario.

Electronic transmission, $\tau$(E),  is calculated by counting the number of contributing channels to the spectrum  at a given energy. This method was shown to be successful in calculating the ballistic TE properties of the 2D transition metal dichalcogenides, group III-VI and group-VA monolayers.\cite{PhysRevB.100.085415,PhysRevB.103.165422,doi:10.1021/acs.nanolett.7b00366} In order to compute $\tauel$ accurately, using a dense k-point sampling is a necessary condition. In our electrical transmission calculations, the number of sampling points was set $101\times101\times1$ for both PBE and HSE06 based calculations. The electrical conductance $G_e=e^2 L_0$, the Seebeck coefficient (thermopower) $S=(L_1/L_0)/eT$, the power factor $PF = S^2G_e$ and the electrons' contribution to the total thermal conductance $\kappa_{el}=(L_2-L_1^{2}/L_0)/T$, were  expressed using the $L_n$ integrals within the Landauer formalism\cite{sivan1986multichannel,esfarjani2006thermoelectric} using 

\begin{equation}
L_n(\mu,T) = -\frac{2}{h}\int \tau(E) (E-\mu)^{n} \left(-\frac{\partial f_{FD}}{\partial E}\right) dE,
\end{equation}

where $\partial f_{FD}(E)/\partial E$ is the derivative of the Fermi-Dirac distribution function.  The partial derivative $\partial f_{FD}(E)/\partial E$ is defined as the Fermi window function, which describes the energy window over conducting channels that contribute to the electronic transport. Similarly, the Landauer formalism is used to obtain lattice thermal conductance, which is expressed as

\begin{equation}
 \kappa_{ph} = \frac{1}{2\pi} \int \tau_{ph}(\omega) \hbar \omega \left(\frac{\partial f_{BE} (\omega, T)}{\partial T}\right)d\omega.\cite{datta1997electronic,rego1998quantized,sevinccli2019green} 
\end{equation}

 Here, $f_{BE} (\omega, T)$ is the Bose-Einstein distribution function and $\tau_{ph}(\omega)$ is the phonon transmission spectrum per width, which is obtained using a $101\times101\times1$ q-point sampling. Thermoelectric figure of merit, $ZT=S^2GT/(\kappa_{el}+\kappa_{ph})$, is calculated using the electronic and lattice thermal transport parameters.

\subparagraph{Acknowledgments:}

The authors thank H\^{a}ldun Sevinçli  for valuable discussions. VOO acknowledges funding from the Scientific and Technological Research Council of Turkey (TUBITAK) under project no 123F264. GÖS acknowledges the support from the TUBITAK project, 123C159.  Computing resources used in this work were provided by the National Center for High Performance Computing of Turkey (UHeM) under grant numbers 1013432022 and 1019342024.

\bibliography{arxiv.bbl}

\end{document}